\documentclass[12pt,oneside,reqno]{amsart}
\usepackage[letterpaper,left=1.40in,right=1.40in,top=1.25in,bottom=1.25in]{geometry}
\usepackage{rotating}
\usepackage[T1]{fontenc}
\usepackage{amsmath}
\usepackage{amsthm,amsfonts}
\usepackage{amssymb,latexsym}
\usepackage{graphicx,rotating}
\usepackage{exscale}
\usepackage{upref}
\usepackage{txfonts}
\usepackage{hyperref}
\usepackage{tabularx} 
\usepackage[normalem]{ulem} 

\usepackage{color}

\theoremstyle{plain}

\theoremstyle{definition}

\theoremstyle{remark}

\numberwithin{equation}{section}

\def\E{\mathrm{I\! E}}

\newcommand{\mat}[1]{\mbox{\boldmath{$#1$}}}

\newcommand{\UP}{{\footnotesize UP}}
\newcommand{\AR}{{\footnotesize AR}}
\newcommand{\FR}{{\footnotesize FR}}
\newcommand{\UC}{{\footnotesize UC}}
\newcommand{\ST}{{\footnotesize ST}}

\newcommand{\SI}{{\footnotesize SI}}

\newcommand{\TO}{{\footnotesize TO}}

\newcommand{\NI}{{\footnotesize NI}}

\newcommand{\STR}{{\footnotesize STR}}

\newcommand{\NB}{{\footnotesize NB}}
\newcommand{\LAT}{{\footnotesize LAT}}
\newcommand{\LON}{{\footnotesize LON}}
\newcommand{\VN}{{\footnotesize VN}}
\newcommand{\FRVN}{{\footnotesize FRVN}}

\newcommand{\SZ}{{\footnotesize SZ}}

\begin{document}

\title[Real estate appraisal of land lots using GAMLSS models]
      {Real estate appraisal of land lots using GAMLSS models}
      
\author{Lutemberg Florencio}
\address{Banco do Nordeste do Brasil S.A\\
         Boa Vista\\
         Recife/PE, 50060--004, Brazil}
\email[L.\ Florencio]{lutemberg@bnb.gov.br}

\author{Francisco Cribari-Neto}
\address{Departamento de Estatística\\
         Universidade Federal de Pernambuco\\
     Cidade Universitária\\
         Recife/PE, 50740--540, Brazil}
\email[F.\ Cribari-Neto]{cribari@de.ufpe.br}
\urladdr[F.\ Cribari-Neto]{\url{http://www.de.ufpe.br/~cribari}}

\author{Raydonal Ospina}
\address{Departamento de Estatística\\
         Universidade Federal de Pernambuco\\
     Cidade Universitária\\
         Recife/PE, 50740--540, Brazil}
\email[R.\ Ospina]{raydonal@de.ufpe.br}
\urladdr[R.\ Ospina]{\url{http://www.de.ufpe.br/~raydonal}}

\keywords{Cubic splines, GAMLSS regression models, hedonic prices function, nonparametric smoothing, semiparametric models. \\
{\it Mathematics Subject Classification: {\rm Primary 62P25 - Secondary 62P20, 62P05}}.
}

\date{\today}

\begin{abstract}
The valuation of real estates is of extreme importance for decision making. Their singular characteristics make valuation through hedonic pricing methods difficult since the theory does not specify the correct regression functional form nor which explanatory variables should be included in the hedonic equation. In this article we perform real estate appraisal using a class of regression models proposed by Rigby \& Stasinopoulos (2005): generalized additive models for location, scale and shape (GAMLSS). Our empirical analysis shows that these models seem to be more appropriate for estimation of the hedonic prices function than the regression models currently used to that end.  
\end{abstract}

\maketitle

\section{Introduction}\label{S:1}

The real estate, apart from being a consumer good that provides comfort and social status, is one of the economic pillars of all modern societies. It has become a form of stock capital, given the expectations of increasing prices, and a means of obtaining financial gains through rental revenues and sale profits. As a consequence, the real estate market value has become a parameter of extreme importance. 

The estimation of a real estate value is usually done using a hedonic pricing equation according to the methodology proposed by Rosen (1974). It is seen as a heterogeneous good comprised of a set of characteristics and it is then important to estimate an explicit function, called hedonic price function, that determines which are the most influential attributes, or attribute `package', when it comes to determing its price. However, the estimation of a hedonic equation is not a trivial task since the theory does not determine the exact functional form nor the relevant conditioning variables.

The use of classical regression methodologies, such as the classical normal linear regression model (CNLRM), for real estate appraisal can lead to biased, inefficient and/or inconsistent estimates given the inherent characteristics of the data (e.g., non-normality, heteroskedasticity and spatial correlation). The use of generalized linear models (GLM) is also subject to shortcomings, since the data may come from a distribution outside the exponential family and the functional relationship between the response and some conditioning variables may not be the same for all observations. There are semiparametric and nonparametric hedonic price estimations available in the literature, such as Pace (1993), Anglin \& Gencay (1996), Gencay \& Yang (1996), Thorsnes \& McMillen (1998), Iwata et al.\ (2000) and Clapp et al.\ (2002). We also highlight the work of Martins-Filho \& Bin (2005), who modeled data from the real estate market in Multnomah County (Oregon-USA) nonparametrically. We note that the use of nonparametric estimation strategies require very large datasets in order to avoid the `curse of dimensionality'. Overall, however, most hedonic price estimations are based on traditional methodologies such as the classical linear regression model and the class of generalized linear models. 

This article proposes a methodology for real estate mass appraisal\footnote{Evaluation of a set of real properties through methodology and procedures common to all of them.} based on the class of GAMLSS models. The superiority GAMLSS modeling relative to traditional methodologies is evidenced by an empirical analysis that employs data on urban land lots located in the city of Aracaju, Brazil. We perform a real estate evaluation in which the response variable is the unit price of land lots and the independent variables are reflect the land lots structural, locational and economic characteristics. We estimate the location and scale effects semiparametrically in such a way that some covariates (the ge\-o\-graph\-ic\-al coordinates of the land lot, for instance) enter the predictor nonparametrically and their effects are estimated using smoothing splines\footnote{For more details on smoothing splines, see Silverman (1984) and Eubank (1999).} whereas other regressors are included in the predictor in the usual parametric fashion. The model delivers a fit that is clearly superior to those obtained using the usual approaches. In particular, we note that our fit yields a very high pseudo-$R^2$. 

The paper unfolds as follows. In Section 2, we briefly present the class of GAMLSS models and highlight its main advantages. In Section 3, we describe the data used in the empirical analysis. In Section 4, we present and discuss the empirical results. Finally, Section 5 closes the paper with some concluding remarks. 
 
\section{GAMLSS modeling}\label{S:2}

\subsection{Definition}

Rigby \& Stasinopoulos (2005) introduced a general class of statistical models called `additive models for location, scale and shape' (GAMLSS). It encompasses both parametric and semiparametric models, and includes a wide range of continuous and discrete distributions for the response variable. It also allows the symultaneous modeling of several parameters that index the response distribution using parametric and/or nonparametric functions. With GAMLSS models, the distribution of the response variable is not restricted to the exponential family and different additive terms can be included in the regression predictors for the parameters that index the distribution, like smoothing splines and random effects, which yields extra flexibility to the model. The model is parametric in the sense that the specification of a distribution for the response variable is required and at the same time it is  semiparametric because one can model some conditioning effects through nonparametric functions.

The probability density function of the response variable $y$ shall be denoted as $f\:\!(y|{\mbox{\boldmath{$\theta$)}}}$, where ${\mbox{\boldmath{$\theta$}}}=(\theta_{1}, \theta_{2},\ldots, \theta_{p})^{\top}$ is a $p$-dimensional parameter vector. It is assumed to belong to a wide class of distributions that we denote by $\mathcal{D}$. This class of distributions includes continuous and discrete distributions as well as truncated, censored and finite mixtures of distributions. In the GAMLSS regression framework the $p$ parameters that index $f(y|{\mbox{\boldmath{$\theta$)}}}$ are modeled using additive terms.

Let ${\mbox{\boldmath{$\tt{y}$}}} = (y_{1}, y_{2},\ldots, y_{n})^{\top}$ be the vector of independent observations on the response variable, each $y_i$ having probability density function $f\:\!(y_i|{\mbox{\boldmath{$\theta^i$)}}}$,
$i=1,\ldots ,n$. Here, ${\mbox{\boldmath{$\theta^{i}$}}} = (\theta_{i1}, \theta_{i2},\ldots, \theta_{ip})^{\top}$ is a vector of $p$ parameters associated to the explanatory variables and to random effects. When the covariates are stochastic, $f(y_i|{\mbox{\boldmath{$\theta^{i}$)}}}$ is taken to be conditional on their values. Additionally, for  $k = 1, 2,\ldots, p$, $g_{k}(\cdot)$ is a strictly monotonic link function that relates the $k$th parameter ${\mbox{\boldmath{$\theta$}}}_{k}$ to explanatory variables and random effects through an additive predictor: 
\begin{equation}\label{eq:1}
g_{k}({\mbox{\boldmath{$\theta$}}}_{k}) = {\mbox{\boldmath{$\eta$}}}_{k} = {\mbox{\boldmath{$X$}}}_{k}\mat{\beta}_{k} + \sum^{J_{k}}_{j=1}{\mbox{\boldmath{$Z$}}}_{jk}\mat{\gamma}_{jk},
\end{equation}
where {\mbox{\boldmath{$\theta$}}}$_{k}$ and ${\mbox{\boldmath{$\eta$}}}_{k}$ are $n\times 1$ vectors,
$\mat{\beta}_{k}=(\beta_{1k}, \beta_{2k},\ldots,\beta_{J'_{k}k})^{\top}$ is a vector of parameters of length $J'_{k}$ and {\mbox{\boldmath{$X$}}}$_{k}$ and {\mbox{\boldmath{$Z$}}}$_{jk}$ are fixed (covariate) design matrixes of orders $n \times J'_{k}$ and $n \times q_{jk}$, respectively. Finally, $\mat{\gamma}_{jk}$ is a $q_{jk}$-dimensional random variable. Model (\ref{eq:1}) is called GAMLSS (Rigby \& Stasinopoulos, 2005). 

In many practical situations it suffices to model four parameters ($p=4$), usually location $(\theta_{1}=\mu)$, scale $(\theta_{2}=\sigma)$, skewness $(\theta_{3}=\nu)$ and kurtosis $(\theta_{4}=\tau)$; the latter two are said to be shape parameters. We thus have the following model: 
\begin{eqnarray}\label{eq:6} \left. \begin{array}{l}
\begin{array}{l}
\mbox{Location and scale} \\
{\mbox {parameters}} 
 \end{array}
 
\left\{ \begin{array}{l} g_{1}({\mbox{\boldmath{$\mu$}}}) = {\mbox{\boldmath{$\eta$}}}_{1} = {\mbox{\boldmath{$X$}}}_{1}\mat{\beta}_{1} + \sum^{J_{1}}_{j=1}{\mbox{\boldmath{$Z$}}}_{j1}\mat{\gamma}_{j1}, \\\\ 
g_{2}({\mbox{\boldmath{$\sigma$}}}) = {\mbox{\boldmath{$\eta$}}}_{2} = {\mbox{\boldmath{$X$}}}_{2}\mat{\beta}_{2} + \sum^{J_{2}}_{j=1}{\mbox{\boldmath{$Z$}}}_{j2}\mat{\gamma}_{j2}, 
\end{array}  \right. \\\\

\mbox{Shape parameters} \quad\;\, \left\{ \begin{array}{l}
g_{3}({\mbox{\boldmath{$\nu$}}}) = {\mbox{\boldmath{$\eta$}}}_{3} = {\mbox{\boldmath{$X$}}}_{3}\mat{\beta}_{3} + \sum^{J_{3}}_{j=1}{\mbox{\boldmath{$Z$}}}_{j3}\mat{\gamma}_{j3},\\\\
g_{4}({\mbox{\boldmath{$\tau$}}}) = {\mbox{\boldmath{$\eta$}}}_{4} = {\mbox{\boldmath{$X$}}}_{4}\mat{\beta}_{4} + \sum^{J_{4}}_{j=1}{\mbox{\boldmath{$Z$}}}_{j4}\mat{\gamma}_{j4}. 

\end{array}  \right.

\end{array}  \right\}  
\end{eqnarray} 
\vskip0.3cm
It is also possible to add to the predictor functions $h_{jk}$ that involve smoothers like cubic splines, penalized splines, fractional polynomials, loess curves, terms of variable coefficients, and others. Any combination of these functions can be included in the submodels for $\mu$, $\sigma$, $\nu$ and $\tau$. As Akantziliotou et al.~(2002) point out, the GAMLSS framework can be applied to the parameters of any population distribution and generalized to allow the modeling of more than four parameters.

GAMLSS models can be estimated using the \texttt{gamlss} package for \textsc{R} (Ihaka \& Gentleman, 1996; Cribari-Neto \& Zarkos, 1999), which is free software; see \url{http://www.R-project.org}. Practitioners can then choose from more than 50 response distributions.

\subsection{Estimation}

Two aspects are central to the GAMLSS additive components fitting, namely: the backfitting algorithm and the fact that quadratic penalties in the likelihood function follow from the assumption that all random effects in the linear predictor are normally distributed.

Suppose that the random effects $\mat{\gamma}_{jk}$ in Model (\ref{eq:1}) are independent and normally distributed with $\gamma_{jk} \sim N_{q_{jk}} ({\mbox{\boldmath{$0$}}},\;\!{\mbox{\boldmath{$G$}}}^{-1}_{jk})$, where ${\mbox{\boldmath{$G$}}}^{-1}_{jk}$ is the $q_{jk} \times q_{jk}$ (generalized) inverse of the symmetric matrix ${\mbox{\boldmath{$G$}}}_{jk} = {\mbox{\boldmath{$G$}}}_{jk}({\mbox{\boldmath{$\lambda$}}}_{jk})$. Rigby \& Stasinopoulos (2005) note that for fixed values of ${\mbox{\boldmath{$\lambda$}}}_{jk}$, one can estimate ${\mbox{\boldmath{$\beta$}}}_{k}$ and ${\mbox{\boldmath{$\gamma$}}}_{jk}$ by maximizing the following penalized log-likelihood function: 
\begin{equation}\label{eq:7}
\ell_{p} = \ell - \frac{1}{2}\sum^{p}_{k=1}\sum^{J_{k}}_{j=1}{\mbox{\boldmath{$\gamma$}}}^{\top}_{jk}{\mbox{\boldmath{$G$}}}_{jk}{\mbox{\boldmath{$\gamma$}}}_{jk},
\end{equation}
where $\ell = \sum^{n}_{i=1}\log\{f(y_{i}|{\mbox{\boldmath{$\theta$}}}^{i})\}$ is the log-likelihood function of the data given ${\mbox{\boldmath{$\theta$}}}^{i}$, for $i = 1, 2, \ldots, n$. This can be accomplished by using a backfitting algorithm.$\footnote{For details, see Rigby \& Stasinopoulos~(2005, 2007), Hastie \& Tibshirani~(1990) and Härdle et al.~(2004).}$

\subsection{Model selection and diagnostic}

GAMLSS model selection is performed by comparing various competing models in which different combinations of the components $\mathcal{M}=\{\mathcal{D}, \mathcal{G}, \mathcal{T}, {\mbox{\boldmath{$\lambda$}}}\}$ are used, where $\mathcal{D}$ specifies the distribution of the response variable, $\mathcal{G}$ is the set of link functions ($g_{1}, \ldots, g_{p}$) for the parameters $(\theta_{1}, \ldots, \theta_{p})$, $\mathcal{T}$ defines the set of predictor terms ($t_{1}, \ldots, t_{p}$) for the predictors $(\eta_{1}, \ldots, \eta_{p})$ and {\mbox{\boldmath{$\lambda$}}} specifies  the set of hiperparameters.

In the parametric GAMLSS regression setting, each nested model $\mathcal{M}$ can be assessed from its fitted global deviance (GD), given by $\mathrm{GD}=-2\ell(\,\hat{\!\mat{\theta}})$, where  $\ell(\,\hat{\!\mat{\theta}})=\sum^{n}_{i=1}\ell(\:\hat{\mat{\!\theta}}^{i})$. Two nested competing GAMLSS models $\mathcal{M}_{0}$ and $\mathcal{M}_{1}$, with fitted global deviances $\mathrm{GD}_{0}$ and $\mathrm{GD}_{1}$ and error degrees of freedom $df_{e0}$ and $df_{e1}$, respectively, can be compared using the generalized likelihood ratio test statistic $\Lambda= \mathrm{GD}_{0} - \mathrm{GD}_{1}$, which is asymptotically distributed as $\chi^{2}$ with $d= {df}_{e0} - {df}_{e1}$ degrees of freedom under $\mathcal{M}_{0}$. For each model $\mathcal{M}$ the number of error degrees of freedom $df_{e}$ is $df_{e} = n - \sum^{p}_{k=1}df_{\theta k}$, where $df_{\theta k}$ are the degrees of freedom that are used in the predictor of the model for the parameter $\theta_{k}$, $k=1,\ldots,p$.

When comparing non-nested GAMLSS models (including models with smoothing terms), the generalized Akaike information criterion (GAIC; Akaike, 1983) can be used to penalize overfittings. That is achieved by adding to the fitted global deviances a fixed penalty $\#$ for each effective degree of freedom that is used in the model, that is, $\mathrm{GAIC} (\#)=\mathrm{GD}+\#df$, where $df$ denotes the total effective number of degrees of freedom that are used in the model and $\mathrm{GD}$ is the fitted global deviance. One then selects the model with the smallest $\mathrm{GAIC}(\#)$ value.

To assess the overall adequacy of the fitted model, we propose the randomized quantile residual (Dunn \& Smyth, 1996). It is a randomized version of the 
Cox \& Snell~(1968) residual and given by
\begin{equation*}
\label{residalet}
r^q_i = \Phi^{-1}(u_i), \quad i=1,\ldots,n,
\end{equation*} 
where $\Phi(\cdot)$ denotes the standard normal distribution function, $u_i$ is a uniform random variable on the interval $(a_i, b_i],$ with $a_i= \lim_{y\uparrow y_i}F\:\!(y_i|{\mbox{\boldmath{$\theta^i$)}}}$ and
$b_i = F\:\!(y_i|{\mbox{\boldmath{$\theta^i$)}}}$. 
A plot of these residuals against the index of the observations $(i)$ should show no detectable pattern. A detectable trend in the plot of some residual against the predictors may be suggestive of link function misspecification. 

Also, normal probability plots with simulated envelopes (Atkinson,~1985) or Worm plots (Buuren \& Fredriks,~2001) are a helpful diagnostic tool. The Worm plots were first introduced by Buuren \& Fredriks (2001) and are useful for analyzing the residuals in different regions (intervals) of the explanatory variable. If no explanatory variable is specified, the worm plot becomes a detrended normal QQ plot of the (normalized quantile) residuals. When all points lie inside the (dotted) confidence bands (the two elliptical curves) there is no evidence of model mispecification.

In the context of a fully parametric model GAMLSS we can use pseudo $R^2$ measures. For example, $R^{2*}_p=1-\log \widehat L/\log \widehat L_0$ 
(McFadden,~1974) and $R^2_{{\rm LR}} = 1-(\widehat{L_0}/\widehat L)^{2/n}$ 
(Cox \& Snell,~1989, pp.\ 208-209), where $\widehat L_0$ and $\widehat L$ are the maximized 
likelihood functions of the null (intercept only) and fitted (unrestricted) models, respectively.  
The ratio of the likelihoods or log-likelihoods may be regarded as a measure of the improvement,
over the model with  ${\mbox{\boldmath{$\theta^i$}}}$ parameters achieved by the model under investigation. 

Our proposal, however, is to compare the different models using the pseudo-$R^2$ given by the square of the sample correlation coefficient between the response and the fitted values. Notice that by doing so we can consider both fully parametric models and models that include nonparametric components. We can also compare the explanatory power of a GAMLSS model to those of GLM and CNLRM models. This is the pseudo-$R^2$ we shall use. It was introduced by Ferrari and Cribari-Neto~(2004) in the context of beta regressions and it is a straightfoward generalization of the $R^2$ measure used in linear regression analysis.

\section{Data description}\label{S:3}

The data contain 2,109 observations on empty urban land lots located in the city of Aracaju, capital of the state of Sergipe (SE), Brazil, and comes from two sources: (i) data collected by the authors from real estate agencies, advertisements on newspapers and research on location (land lots for sale or already sold);  (ii) data obtained from the `Departamento de Cadastro Imobili\'{a}rio da Prefeitura de Aracaju'. Observations cover the following years: 2005, 2006 and 2007. Each land lot price was recorded only once during that period. It is also noteworthy that the land lots in the sample are geographically referenced relative to the South American Datum$\footnote{The South American Datum (SAD) is the regional geodesic system for South America and refers to the mathematical representation of the Earth surface at sea level.}$ and have their geographical positions (latitude, longitude) projected onto the Universal Transverse Mercator (UTM) coordinate system.$\footnote{Cilindrical cartographic projection of the terrestrial spheroid in 60 secant cylinders at Earth level alongside the meridians in multiple zones of 6 degrees longitude and stretching out 80 degrees South latitude to 84 degrees North latitude.}$

The sample used to estimate the hedonic prices equation\footnote{That is, the equation of hedonic prices of urban land lots in Aracaju-SE.} contains, besides the year of reference, information on the physical (area, front, topography, pavement and block position), locational (neighborhood, geographical coordinates, utilization coefficient and type of street in which the land lot is located) and economic (nature of the information that generated the observation, average income of the head of household of the censitary system where the land is located and the land lot price) characteristics of the land lots. In particular, we shall use the following variables: 
\begin{itemize}
	  \item {\footnotesize YEAR (YR)}: qualitative ordinal variable that identifies the year in which the information was obtained. It assumes the values 2005, 2006 (YR06) and 2007 (YR07). It enters the model through dummy variables; 
	  \vskip0.1cm
		\item {\footnotesize AREA (AR)}: continuous quantitative variable, measured in $\rm{m}^{2}$ (square meters), relative to the projection on a horizontal plane of the land surface; 
		\vskip0.1cm
		\item {\footnotesize FRONT (FR)}: continuous quantitative variable, measured in \rm{m} (meters), con\-cer\-ning the projection of the land lot front over a line which is perpendicular to one of the lot boundaries, when both are oblique in the same sense, or to the `chord', in the case of curved fronts;
		\vskip0.1cm
		\item {\footnotesize TOPOGRAPHY (TO)}: nominal qualitative variable that relates to the topographical conformations of the land lot. It is classified as `plain' when the land acclivity is smaller than $10\%$ or its declivity is smaller than $5\%$, and as `rough' otherwise. It is a dummy variable that equals 1 for `plain' and 0 `rough'; 
		\vskip0.1cm
		\item {\footnotesize PAVEMENT (PA)}: nominal qualitative variable that indicates the presence or absence of pavement (concrete, asphalt, among others) on the street in which the main land lot front is located. It enters the model as a dummy variable that equals 1 when the land lot is located on a paved street and 0 otherwise;
		\vskip0.1cm
		\item {\footnotesize SITUATION (SI)}: nominal qualitative variable used to differentiate the disposition of the land lot on the block. It is classified as `corner lot' or `middle lot'. It is a dummy variable that assumes value 1 for corner lots and 0 for all other land lots;
		\vskip0.1cm
		\item {\footnotesize NEIGHBORHOOD (NB)}: nominal qualitative variable referring to the name of the neigh\-bor\-hood where the land lot is located. It was categorized as valuable (highly priced) neighborhoods and other neighborhoods, with the variable shown as \VN\ and regarded as a dummy (1 for valuable neighborhoods). The neighborhoods were also grouped as belonging or not belonging to the city South Zone, dummy denoted by \SZ\ (1 for South Zone);  
		\vskip0.1cm
	\item {\footnotesize LATITUDE (LAT) and LONGITUDE (LON)}: continuous quantitative variables corresponding to the geographical position of the land lot at the point $z =$ ({\footnotesize LAT, LON}), where {\footnotesize LAT} and {\footnotesize LON} are the coordinates measured in UTM;
	\vskip0.1cm
	\item {\footnotesize UTILIZATION COEFFICIENT (UC)}: discrete variable given by a number that, when multiplied by the area of the land lot, yields the maximal area (in square meters) available for construction. \UC\ is defined in an official urban development document. It assumes the following values: $3.0, 3.5,\ldots,5.5, 6.0$;
	\vskip0.1cm
	\item {\footnotesize STREET (STR)}: ordinal qualitative variable used to differentiate the land lot location relative to streets and avenues. It is classified as `minor arterial' (STR1), `collector street' (STR2) and `local street' according to the importance of the street where the land lot is located. It enters the model as dummy variables; 
	\vskip0.1cm
		\item {\footnotesize NATURE OF THE INFORMATION (NI)}: nominal qualitative variable that indicates whether the observation is derived from `offer', `transaction' or from the Aracaju register office (real state sale taxes). It enters the model through dummy variables; 
		\vskip0.1cm
		\item {\footnotesize SECTOR (ST)}: discrete quantitative proxy variable of macrolocation used to socioeconomically distinguish the various neighborhoods, represented by the average income of the head of household, in minimum wages, according to the IBGE census (2000). The neighborhood average income functions as a proxy to other characteristics, such as urban amenities.  It assumes the following values:  $1,2,\ldots,18$; 
		\vskip0.1cm
		\item {\footnotesize FRONT IN HIGHLY VALUED NEIGHBORHOODS (FRVN)}: continuous quantitative variable that assumes strictly positive values and corresponds to the interaction between \FR\ and \VN\ variables. It is included in the model to capture the influence of land lots front dimensions in `valuable' neighborhoods;
		\vskip0.1cm
		\item {\footnotesize UNIT PRICE (UP)}: continuous quantitative variable that assumes strictly positive values and corresponds to the land lot price divided by its area, measured in $\rm{R}\$/\rm{m}^{2}$ (reais per square meter).
	 \end{itemize} 

In real estate appraisals (and specifically in land lots valuations), the interest typically lies in modeling the unit price as a function of the underlying structural, locational and economic characteristics of the real estate. We shall then use \UP\ as the dependent variable (response). The independent variables relate to the locational (\NB, \VN, \SZ, \LAT, \LON, \ST, \UC\ and \STR), physical (\AR, \FR, \TO, \SI\ and \FRVN) and economic (\NI) land lot characteristics; we also account for the year of data collection. 

\begin{figure}[!htb]
	\centering
		\includegraphics[height=6.7cm,width=12cm]{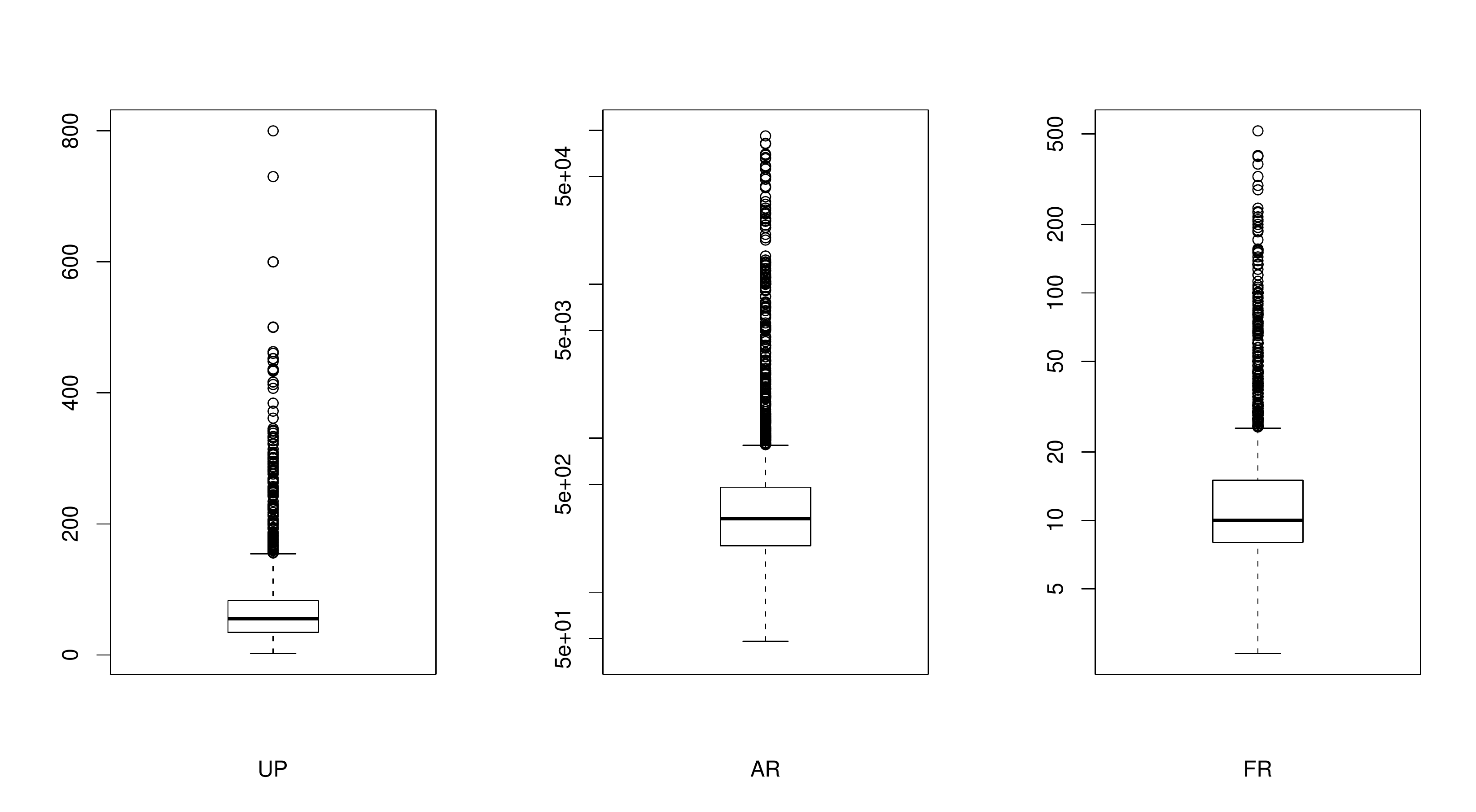}
		\vskip -0.5cm
	\caption{Box-plots of \UP, \AR\ and \FR.}	\label{figbpPUARFR}
\end{figure}

Figure~\ref{figbpPUARFR} presents box-plots of \UP, \AR\ and \FR\ and Table~\ref{tabVARQUANT} displays summary statistics on those variables. The box-plot of \UP\ shows that its distribution is skewed and that there are several extreme observations. Notice from Table~\ref{tabVARQUANT} that the sample values of \UP\ range from $\rm{R}\$\: 2.36/\rm{m}^2$ to $\rm{R}\$\:\,800.00/\rm{m}^2$ and that $75\%$ of the land lots have unit prices smaller than $\rm{R}\$\:82.82/\rm{m}^2$.

We note that 263 extreme observations have been identified from the box-plot of \AR\ (see Figure~\ref{figbpPUARFR}). These observations are not in error, they appear as outlying data points in the plot because the variable assumes a quite wide range of values: from $41\: \rm{m}^2$ to $91,780\: \rm{m}^2$, that is, the largest land lot is nearly two thousand times larger than the smallest one.

\begin{table}[!ht]
\begin{center}
\caption{Descriptive statistics.}\label{tabVARQUANT}
\vskip 0.0cm
\tiny
\begin{tabular}{rrrrrrr}
  \hline
Variable & Mean & Median & Standard error & Minimum & Maximum & Range \\ 
  \hline
UP & 72.82 & 55.56 & 70.28 & 2.36 & 800.00 & 797.64 \\ 
LAT & 710100.00 & 710300.00 & 2722.34 & 701500.00 & 714600.00 & 13100.00 \\ 
LON & 8787000.00 & 8786000.00 & 6638.77 & 8769000.00 & 8798000.00 & 29000.00 \\ 
AR & 1355.00 & 300.00 & 6063.53 & 48.00 & 91780.00 & 91732.00 \\ 
FR & 18.13 & 10.00 & 30.54 & 2.60 & 516.00 & 513.40 \\ 
   \hline
\end{tabular}
\end{center}
\end{table}

\begin{figure}[!ht]
	\centering
		\includegraphics[width=\linewidth]{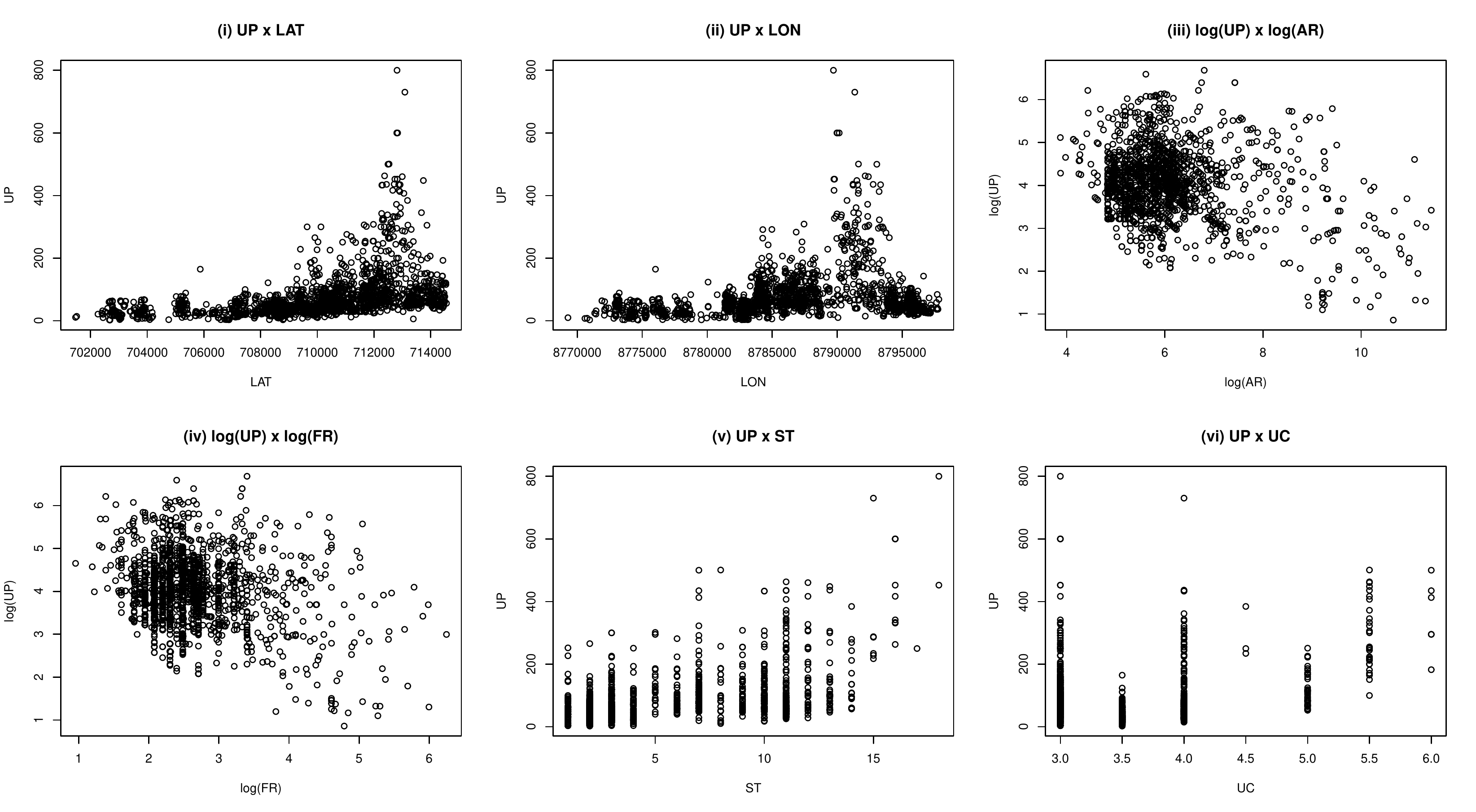}
	\caption{Dispersion plots.}	\label{figddPUDEMAIS}
\vskip-0.2cm
\end{figure}

In order to investigate how \UP\ relates to some explanatory variables, we produced dispersion plots. Figure~\ref{figddPUDEMAIS} contains the following pairwise plots: (i) \UP\ $\times$ \LAT; (ii) \UP\ $\times$ \LON; (iii) log(\UP) $\times$ log(\AR); (iv) log(\UP) $\times$ log(\FR); (v) \UP\ $\times$ \ST\ and (vi) \UP\ $\times$ \UC. 
It shows that there is a direct relationship between \UP\ and the corresponding regressor in (i), (ii), (v) and (vi), whereas in (iii) and (iv) the relationship is inverse. Thus, there is a tendency for the land lot unit price to increase with latitude, longitude, sector and also with the utilization coefficient, and to decrease as the area and the front size increase. We note that the inverse relationship between unit price and front size was not expected. It motivated the inclusion of the covariate \FRVN\ in our analysis. 

It is not clear from Figure~\ref{figddPUDEMAIS} whether the usual assumptions of normality and homoskedasticity are reasonable. As noted by Rigby \& Stasinopoulos (2007), transformations of the response variable and/or of the explanatory variables are usually made in order to minimize deviations from the underlying assumptions. However, this practice may not deliver the expected results. Additionally, the resulting model parameters are not typically easily interpretable in terms of the untransformed variables. A more general modeling strategy is thus called for.

\section{Empirical modeling}\label{S:4}

In what follows we shall estimate the hedonic price function of land lots located in Aracaju using the highly flexible class of GAMLSS models. At the outset, however, we shall estimate standard linear regression and generalized linear models. We shall use these fits as benchmarks for our estimated GAMLSS hedonic price function.

\subsection{Data modeling based on the CNLRM}

Table~\ref{ajustesCNLRM} lists the classical normal linear regressions that were estimated. The transformation parameter of the Box-Cox model was estimated by maximizing the profile log-likelihood function: $\hat{\!\lambda}=0.1010$. All four models are heteroskedastic and there is strong evidence of nonnormality for the first two models. The coefficients of determination range from 0.54 to 0.66. Since the error variances are not constant, we present in Table~\ref{ajusteCNLRM} the estimated parameters of Model (1.4), which yields the best fit, along with heteroskedasticity-robust HC3 standard errors (Davidson \& MacKinnon, 1993). Notice that all covariates are statistically significant at the 5\% nominal level, except for \LAT\ ($p\textrm{-value}=0.1263$), which suggests that pricing differentiation mostly takes place as we move in the North-South direction.

\newcolumntype{Y}{>{\raggedright\arraybackslash}X}
\newcolumntype{Z}{>{\centering\arraybackslash}X}
\begin{table}[!ht]
\scriptsize
\begin{center}
\caption{Fitted models via CNLRM.}\label{ajustesCNLRM}
\begin{tabularx}{\textwidth}{cXXX}
  \hline
\textrm{Model}& \textrm{Equation}& \textrm{Considerations}\\
\hline\hline
$1.1$ & $\scriptsize  \textrm{UP} = \beta_0+\beta_1\textrm{LAT}+\beta_2\textrm{LON} +\beta_3\textrm{AR}+\beta_4\textrm{UC}+\beta_5\textrm{ST}+\beta_6\textrm{STR1}+\beta_7\textrm{STR2}+\beta_8\textrm{SI}+\beta_9\textrm{PA}+\beta_{10}\textrm{TO}+\beta_{11}\textrm{NIO}+\beta_{12}\textrm{NIT}+\beta_{13}\textrm{YR06}+\beta_{14}\textrm{YR07}+\beta_{15}\textrm{SZ}+\beta_{16}\textrm{FRVN}+\epsilon$ & The null hypotheses that the errors are homoskedastic and normal are rejected at the 1\% nominal level by the Breusch-Pagan and Jarque-Bera tests, respectively. The explanatory variables proved to be statistically significant at the 1\% nominal level ($z$-tests). Also, $\overline{R}^2=0.539$, $\mathrm{AIC}=22304$ and $\mathrm{BIC}=22406$.\\
   \hline
$1.2$ & $\scriptsize \textrm{log(UP)}= \beta_0+\beta_1\textrm{LAT}+\beta_2\textrm{LON}+\beta_3\textrm{AR}+\beta_4\textrm{UC}+\beta_5\textrm{ST}+\beta_6\textrm{STR1}+\beta_7\textrm{STR2}+\beta_8\textrm{SI}+\beta_9\textrm{PA}+\beta_{10}\textrm{TO}+\beta_{11}\textrm{NIO}+\beta_{12}\textrm{NIT}+\beta_{13}\textrm{YR06}+\beta_{14}\textrm{YR07}+\beta_{15}\textrm{SZ}+\beta_{16}\textrm{FRVN}+\epsilon$ & The null hypotheses that the errors are homoskedastic and normal are rejected at the 1\% nominal level by the Breusch-Pagan and Jarque-Bera tests, respectively. All explanatory variables proved to be statistically significant at 1\% the nominal level ($z$-tests). Also, $\overline{R}^2=0.599$, $\mathrm{AIC}=2912$ and $\mathrm{BIC}=3014$.\\
  \hline   
$1.3$ & $\scriptsize \textrm{log(UP)}= \beta_0+\beta_1\textrm{LAT}+\beta_2\textrm{LON}+\beta_3\textrm{log(AR)}+\beta_4\textrm{UC}+\beta_5\textrm{log(ST)}+\beta_6\textrm{STR1}+\beta_7\textrm{STR2}+\beta_8\textrm{SI}+\beta_9\textrm{PA}+\beta_{10}\textrm{TO}+\beta_{11}\textrm{NIO}+\beta_{12}\textrm{NIT}+\beta_{13}\textrm{YR06}+\beta_{14}\textrm{YR07}+\beta_{15}\textrm{SZ}+\beta_{16}\textrm{log(FRVN)}+\epsilon$ & The Jarque-Bera test does not reject the null hypothesis of normality at the usual nominal levels, but the Breusch-Pagan test rejects the null hypothesis of homoskedasticity at the 1\% nominal level. All explanatory variables are statistically significant at the 1\% nominal level, except for the {\tiny LAT} variable ($p\textrm{-value} = 0.0190$). Also, $\overline{R}^2=0.651$, $\mathrm{AIC}=2619$ and $\mathrm{BIC}=2721$.\\
 \hline
$1.4$ & $\scriptsize \frac{\textrm{UP}^{\lambda}-1}{\lambda} = \beta_0+\beta_1\textrm{LAT}+\beta_2\textrm{LON}+\beta_3\textrm{log(AR)}+\beta_4\textrm{UC}+\beta_5\textrm{log(ST)}+\beta_6\textrm{STR1}+\beta_7\textrm{STR2}+\beta_8\textrm{PA}+\beta_9\textrm{TO}+\beta_{10}\textrm{NIO}+\beta_{11}\textrm{NIT}+\beta_{12}\textrm{YR06}+\beta_{13}\textrm{YR07}+\beta_{14}\textrm{log(FRVN)}+\epsilon$ & Normality is not rejected by the Jarque-Bera test, but the Breusch-Pagan test rejects the null hypothesis of homoskedasticity at the 1\% nominal level. All covariates proved to be statistically significant at the 1\% nominal level, except for the {\tiny LAT} variable ($p\textrm{-value} = 0.0881$). Also, $\overline{R}^2=0.657$, $\mathrm{AIC}=4290$ and $\mathrm{BIC}=4392$.\\
   \hline
 \end{tabularx}
 \end{center}
 \end{table}

\begin{table}[!htb]
\scriptsize
\begin{center}
\caption{Hedonic price function estimated via CNLRM -- Model (1.4).}\label{ajusteCNLRM}
\begin{tabular}{rrrrr}
  \hline
 & Estimate & Standard error & $z$-statistic  & $p$-value \\ 
  \hline
(Intercept) & $-162.6307$ & 34.1920 & $-4.756$ & 0.0000 \\ 
  LAT & 1.85e-05 & 1.21e-05 & 1.529 & 0.1263 \\ 
  LON & 1.74e-05 &  4.60e-06 &  3.798 & 0.0001 \\ 
  log(AR) & $-0.3507$ & 0.0192 &  $-18.236$ & 0.0000 \\ 
  log(ST) & 0.4423 & 0.0332 & 13.297 & 0.0000 \\ 
  UC & 0.2651 & 0.0412 & 6.429 & 0.0000 \\ 
  STR1 & 0.4874 & 0.0717 & 6.789 & 0.0000 \\ 
  STR2 & 0.1678 & 0.0675 & 2.485 & 0.0130 \\ 
  SI & 0.1119 & 0.0405 & 2.757 &  0.0058 \\ 
  PA & 0.3853 & 0.0302 & 12.767 & 0.0000 \\ 
  TO & 0.4905 & 0.0798 & 6.145 & 0.0000 \\ 
  NIO & 0.5994 & 0.0592 & 10.131 & 0.0000 \\ 
  NIT & 0.5111 & 0.0131 & 3.886 & 0.0000 \\ 
  YR06 & 0.2560 & 0.0351 & 7.289 & 0.0000 \\ 
  YR07 & 0.6450 & 0.0345 & 18.645 & 0.0000 \\ 
  SZ & 0.7221 & 0.0474 & 15.239 & 0.0000 \\ 
  log(FRVN) & 1.2041 & 0.0137 & 8.797 & 0.0000 \\ 
   \hline
\end{tabular}
\end{center}
\end{table}

\subsection{Hedonic GLM function}

Table~\ref{ajusteGLM} displays the maximum likelihood fit of the following generalized linear model: 
\begin{align*}
g(UP^*)& = \beta_0+\beta_2\textrm{LON}+\beta_3\textrm{log(AR)}+\beta_4\textrm{UC}+\beta_5\textrm{log(ST)}+\beta_6\textrm{STR1}+\beta_7\textrm{STR2}\\
&+ \beta_8\textrm{SI}+\:\beta_9\textrm{PA}+\beta_{10}\textrm{TO}+\beta_{11}\textrm{NIO}+\beta_{12}\textrm{NIT}+\beta_{13}\textrm{YR06}+\beta_{14}\textrm{YR07}\\
&+\beta_{15}\textrm{SZ}+\beta_{16}\textrm{log(FRVN)},\hskip4.9cm \quad \quad \quad \quad (\textrm{Model 2.1})
\end{align*}
where $UP^{*}=\E(UP)=\mu$, \textit{\UP}\ $\sim$ gamma $(\mu, \sigma)$ and $\eta=\log(\mu)$. We have tried a number of different models, and this one (gamma response and log link) yielded the best fit. We also note that all regressors are statistically significant at the 1\% nominal level, except for  \LAT\ ($p$-value = 0.5295), which is why we dropped this covariate from the model.

\begin{table}[!htb]
\scriptsize
\begin{center}
\caption{Hedonic price function estimated via GLM -- Model (2.1).}\label{ajusteGLM}
\begin{tabular}{rrrrr}
  \hline
 & Estimate & Standard error & $z$-statistic  & $p$-value \\ 
  \hline
(Intercept) & $-151.8019$ & 15.7792 & $-9.620$ & 0.0000 \\ 
  LON & 1.77e-05 & 1.80e-06 & 9.851 & 0.0000 \\     
  log(AR) & $-0.2276$ & 0.0108 & $-21.120$ & 0.0000 \\ 
  UC & 0.1272 & 0.0231 & 5.515 & 0.0000 \\ 
  log(ST) & 0.2880 & 0.0193 & 14.954 & 0.0000 \\ 
  STR1 & 0.3562 & 0.0395 & 9.021 & 0.0000 \\ 
  STR2 & 0.1419 & 0.0408 & 3.482 & 0.0005 \\ 
  SI & 0.0945 & 0.0255 & 3.707 & 0.0002 \\ 
  PA & 0.2324 & 0.0220 & 10.556 & 0.0000 \\ 
  TO & 0.3139 & 0.0503 & 6.236 & 0.0000 \\ 
  NIO & 0.4208 & 0.0348 & 12.087 & 0.0000 \\ 
  NIT & 0.3779 & 0.0642 & 5.884 & 0.0000 \\ 
  YR06 & 0.1947 & 0.0242 & 8.035 & 0.0000 \\ 
  YR07 & 0.4551 & 0.0242 & 18.780 & 0.0000 \\ 
  SZ & 0.4716 & 0.0310 & 15.220 & 0.0000 \\ 
  log(FRVN) & 0.7467 & 0.0622 & 11.997 & 0.0000 \\ 
   \hline
\end{tabular}
\end{center}
\vskip-0.5cm
\end{table}

\subsection{GAMLSS hedonic fit}

\subsubsection{Location parameter modeling ($\mu)$}

Since \UP\ (the response) only assumes positive values, we have considered the following distributions for it: log-normal (LOGNO), inverse Gaussian (IG), Weibull (WEI) and gamma (GA). 
As noted earlier, we use pseudo-$R^2$ given by 
\begin{equation}
\label{psdr}
\textrm{pseudo-}R^2=[\mbox{correlation}\:(\mbox{observed values of \UP, predicted values of \UP})]^2
\end{equation}
to measure the overall goodness-of-fit. 

\newcolumntype{Y}{>{\raggedright\arraybackslash}X}
\newcolumntype{Z}{>{\centering\arraybackslash}X}
\begin{table}[!htb]
\scriptsize
\begin{center}
\caption{Fitted models via GAMLSS.}\label{ajustesGAMLSS}
\begin{tabularx}{\textwidth}{cccXXX}
  \hline
Model& $\mathcal{D}$&  $\mathcal{G}$ & Equation &Considerations\\
\hline\hline
$3.1$ & LOGNO & logarithmic &$\scriptsize \textrm {UP}= \beta_0+\textrm{cs(LAT)}+\textrm{cs(LON)}+\textrm{cs(log(AR))}+\textrm{cs(UC)}+\textrm{cs(ST)}+\beta_1\textrm{STR1}+\beta_2\textrm{STR2}+\beta_3\textrm{SI}+\beta_4\textrm{PA}+\beta_{5}\textrm{TO}+\beta_{6}\textrm{NIO}+\beta_{7}\textrm{NIT}+\beta_{8}\textrm{YR06}+\beta_{9}\textrm{YR07}+\beta_{10}\textrm{SZ}+\textrm{cs(log(FRVN))}$  & All regressors are significant at the level 1\% significance level ($z$-tests). Also, $\mathrm{AIC}=19155$, $\mathrm{BIC}=19359$ and $\mathrm{GD}= 19083$. Pseudo-$R^2$= 0.739.\\
   \hline
   
   $3.2$ & IG & logarithmic&$\scriptsize \textrm{UP}= \beta_0+\textrm{cs(LAT)}+\textrm{cs(LON)}+\textrm{cs(log(AR))}+\textrm{cs(UC)}+\textrm{cs(ST)}+\beta_1\textrm{STR1}+\beta_2\textrm{STR2}+\beta_3\textrm{SI}+\beta_4\textrm{PA}+\beta_{5}\textrm{TO}+\beta_{6}\textrm{NIO}+\beta_{7}\textrm{NIT}+\beta_{8}\textrm{YR06}+\beta_{9}\textrm{YR07}+\beta_{10}\textrm{SZ}+\textrm{cs(log(FRVN))}$  & All regressors are significant at the 1\% significance level ($z$-test). Also, $\mathrm{AIC}=19845$, $\mathrm{BIC}=20048$ and $\mathrm{GD}=19773$. Pseudo-$R^2$= 0.678.\\
   \hline
   
   $3.3$ & WEI & logarithmic&$\scriptsize \textrm{UP}= \beta_0+\textrm{cs(LAT)}+\textrm{cs(LON)}+\textrm{cs(log(AR))}+\textrm{cs(UC)}+\textrm{cs(ST)}+\beta_1\textrm{STR1}+\beta_2\textrm{STR2}+\beta_3\textrm{SI}+\beta_4\textrm{PA}+\beta_{5}\textrm{TO}+\beta_{6}\textrm{NIO}+\beta_{7}\textrm{NIT}+\beta_{8}\textrm{YR06}+\beta_{9}\textrm{YR07}+\beta_{10}\textrm{SZ}+\textrm{cs(log(FRVN))}$  & All regressors proved to be significant at the $1\%$ significance level ($z$-tests). Also, $\mathrm{AIC}=19260$, $\mathrm{BIC}=19463$ and $\mathrm{GD}=19188$. Pseudo-$R^2$= 0.748.\\
   \hline
   
   $3.4$ & GA & logarithmic&$\scriptsize \textrm{UP}= \beta_0+\textrm{cs(LAT)}+\textrm{cs(LON)}+\textrm{cs(log(AR))}+\textrm{cs(UC)}+\textrm{cs(ST)}+\beta_1\textrm{STR1}+\beta_2\textrm{STR2}+\beta_3\textrm{SI}+\beta_4\textrm{PA}+\beta_{5}\textrm{TO}+\beta_{6}\textrm{NIO}+\beta_{7}\textrm{NIT}+\beta_{8}\textrm{YR06}+\beta_{9}\textrm{YR07}+\beta_{10}\textrm{SZ}+\textrm{cs(log(FRVN))}$  & All regressors are significant at the 1\% significance level ($z$-tests). Also, $\mathrm{AIC}=19062$, $\mathrm{BIC}=19337$ and $\mathrm{GD}=19134$. Pseudo-$R^2$= 0.746\\
   \hline
   \end{tabularx}
 \end{center}
 \end{table}

The models listed in Table \ref{ajustesGAMLSS} include smoothing cubic splines (cs) with 3 effective degrees of freedom for the covariates \LAT, \LON, log(\AR), \UC, \ST\ and log(\FRVN). Other smoothers (such as loess and penalized splines), as well as different combinations of $\mathcal{D}$ (such as BCPE, BCCG, LNO, BCT, exGAUSS, among others; see Rigby and Stasinopoulos, 2007) and $\mathcal{G}$ (such as identity, inverse, reciprocal, among others), were considered. However they did not yield superior fits. We also note that Model (3.4) yields the smallest values of the three model selection criteria. Table \ref{ajustemod34} contains the a summary of the model fit.

\begin{table}[!htb]
\scriptsize
\begin{center}
\caption{Hedonic price function estimated via GAMLSS -- Model (3.4).}\label{ajustemod34}
\begin{tabular}{rrrrr}
  \hline
 &Estimative  &Standard error  & $z$-statistic  & $p$-value\\ 
  \hline
(Intercept) & $-165.4000$ & 16.1300 & $-10.251$ & 0.0000 \\ 
  cs(LAT) & 5.17e-05 & 6.22e-06 & 8.307 & 0.0000 \\ 
  cs(LON) & 1.51e-05 & 2.13e-06 & 7.071 & 0.0000 \\ 
  cs(log(AR)) & $-0.2317$ & 0.0096 & $-24.074$ & 0.0000 \\ 
  cs(ST) & 0.0465 & 0.0037 & 12.416 & 0.0000 \\ 
  cs(UC) & 0.1223 & 0.0206 & 5.947 & 0.0000 \\ 
  STR1 & 0.3133 & 0.0349 & 8.963 & 0.0000 \\ 
  STR2 & 0.0926 & 0.0364 & 2.545 & 0.0100 \\ 
  SI & 0.0920 & 0.0227 & 4.054 & 0.0000 \\ 
  PA & 0.1891 & 0.0195 & 9.670 & 0.0000 \\ 
  TO & 0.2662 & 0.0474 & 5.951 & 0.0000 \\ 
  NIO & 0.4135 & 0.0395 & 13.362 & 0.0000 \\ 
  NIT & 0.3485 & 0.0571 & 6.102 & 0.0000 \\ 
  YR06 & 0.1645 & 0.0215 & 7.632 & 0.0000 \\ 
  YR07 & 0.4358 & 0.0215 & 20.235 & 0.0000 \\ 
  cs(log(FRVN)) & 0.6513 & 0.0569 & 11.443 & 0.0000 \\ 
  SZ & 0.3875 & 0.0299 & 12.935 & 0.0000 \\ 
   \hline
\end{tabular}
\end{center}
\end{table}

The use of three effective degrees of freedom in the smoothing functions delivered a good model fit. However, in order to determine whether a different number of effective degrees of freedom delivers superior fit, we used two criteria, namely: the AIC (objective) and visual inspection of the smoothed curves (subjective); visual inspection aimed at avoiding overfitting. We then arrived at Model (3.5). It also uses cubic spline smoothing (cs), but with a different number of effective degrees of freedom (df) in the smoothing functions; see Table~\ref{ajustemod35}. Notice that there was a considerable reduction -- relative to Model (3.4) -- in the AIC, BIC and GD values (18822, 19212 and 18684, respectively) and that there is a better agreement between observed and predicted response values.

\begin{table}[!h]
\scriptsize
\begin{center}
\caption{Hedonic price function estimated via GAMLSS -- Model (3.5).}\label{ajustemod35}
\begin{tabular}{rrrrr}
  \hline
&Estimative  &Standard error  & $z$-statistic  & $p$-value\\   
  \hline
(Interceptt) & $-130.1000$ & 14.8100 & $-8.787$ & 0.0000 \\ 
  cs(LAT, df=10) & 5.92e-05 & 5.71e-06 & 10.354 & 0.0000 \\ 
  cs(LON, df=10) & 1.05e-05 & 1.96e-06  & 5.352 & 0.0000 \\ 
  cs(log(AR), df=10) & $-0.2559$ &  8.83e-03 & $-28.963$ & 0.0000 \\ 
  cs(ST, df=8) & 0.0373 & 3.44e-03 & 10.831 & 0.0000 \\ 
  cs(UC, df=3) & 0.1769 & 0.0188 & 9.370 & 0.0000 \\ 
  STR1 & 0.2571 & 0.0320 & 8.012 & 0.0000 \\ 
  STR2 & 0.0728 & 0.0334 & 2.180 & 0.0293 \\ 
  SI & 0.1029 & 0.0208 & 4.940 & 0.0000 \\ 
  PA & 0.1436 & 0.0179 & 7.999 & 0.0000 \\ 
  TO & 0.1822 & 0.0410 & 4.436 & 0.0000 \\ 
  NIO & 0.4173 & 0.0284 & 14.690 & 0.0000 \\ 
  NIT & 0.3388 & 0.0524 & 6.462 & 0.0000 \\ 
  YR06 & 0.1373 & 0.0198 & 6.941 & 0.0000 \\ 
  YR07 & 0.4190 & 0.0197 & 21.190 & 0.0000 \\ 
  cs(log(FRVN), df=10) & 0.6599 & 0.0522 & 12.630 & 0.0000 \\ 
  SZ & 0.5119 & 0.0275 & 18.613 & 0.0000 \\ 
   \hline
\end{tabular}
\end{center}
\end{table}

\begin{figure}[!htb]
	\centering
		\includegraphics[height=14cm,width=13cm]{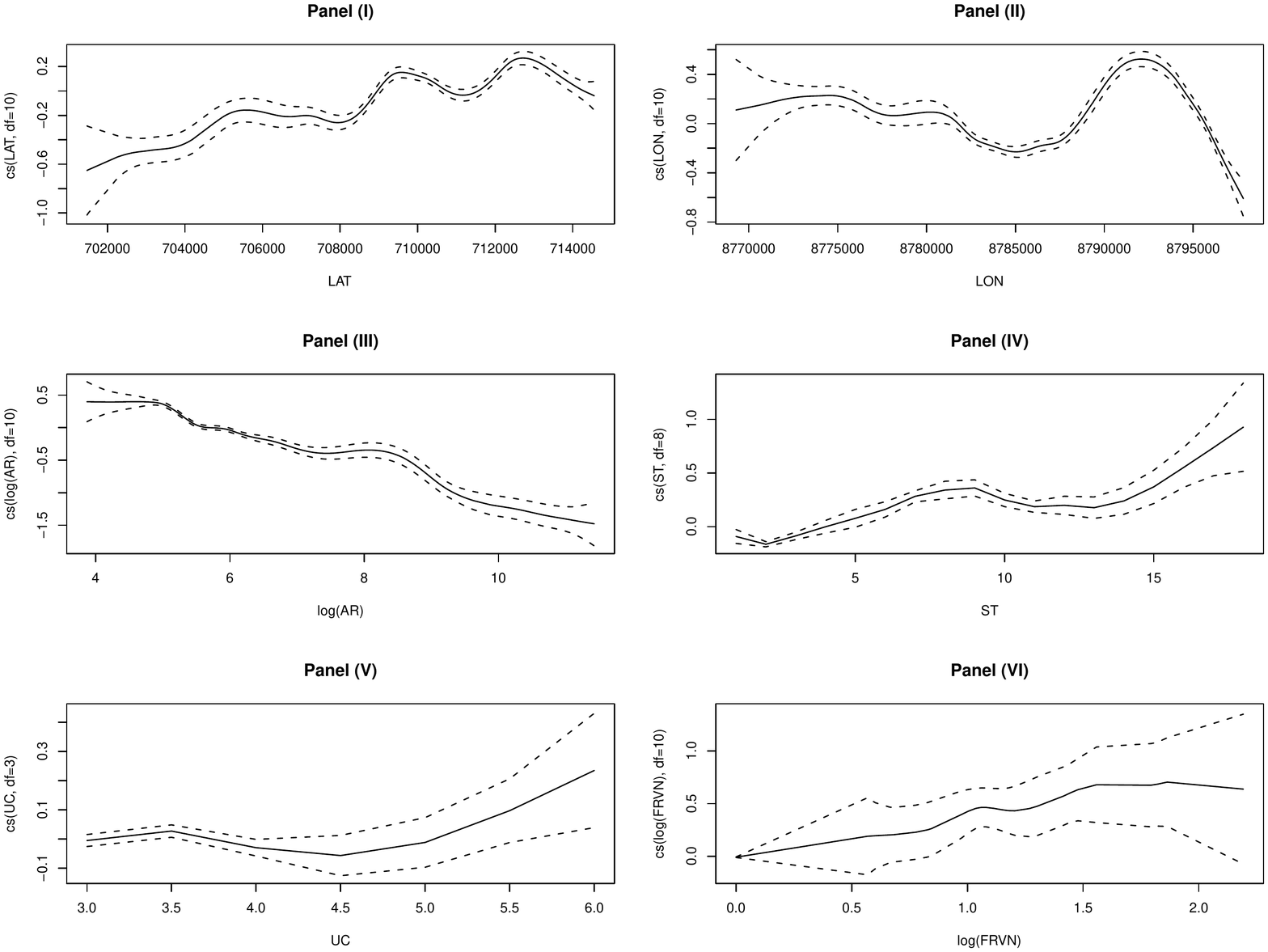}
	\caption{Smoothed additive terms -- Model (3.5).}\label{figtermplotmu}
	\end{figure}

Figure~\ref{figtermplotmu} contains plots of the smoothed curves from Model (3.5). The dashed lines are confidence bands based on pointwise standard errors.
 Panels (I), (II), (III), (IV), (V) and (VI) reveal that the effects/impacts of \LAT, \LON, log(\AR), \ST, \UC\ and log(\FRVN) are typically increasing, increasing/decreasing,\footnote{Panel (II) alternately shows local increasing and decreasing trends.} decreasing, increasing, increasing and increasing, respectively, with increases in latitude, longitude, log area, socioeconomic indicator, utilization coefficient and log land front in highly priced neighborhoods. Some of these effects were also suggested by the estimated coefficients of the CNLRM and GLM models. Here, however, one obtains a somewhat more flexible global picture, as we shall see. 
 
In panel (I), one notices that as the latitude increases the `contribution' of the \LAT\ covariate between the 702000 and 709000 latitudes (approximately) -- neighborhoods that belong to the expansion zone of the city -- is negative, whereas starting from position 709000 (approximately) -- South Zone and downtown area -- the price effect is positive. Additionally, we note that, in certain ranges, increases in latitude lead to drastic changes in the slope of the smoothed curve, e.g., between the 708000 and 710000 positions, whereas in other areas, for instance between the 706000 and 708000 latitudes -- the Mosqueiro neighborhood --, an increase in latitude leads to an uniform negative effect. 

Panel (II) shows that as longitude increases to position 8780000 the  `contribution' of the \LON\ covariate is positive and nearly uniform, which almost exclusively covers observations relative to the Mosqueiro neighborhood. Starting at the 8785000 position there is a remarkable change in the slope of the fitted curve, which is triggered by the location of the most upper class neighborhoods: from 8785000 to 8794000. After the 8794000 position, the effect remains positive, but is decreasing; it eventually becomes negative.

We see in Panel (III) that as the area (in logs) increases the `contribution' of the log(\AR) covariate, for land lots with log areas between 4 and 5 (respectively), is clearly positive. The effect is negative for land lots with log areas in excess of 5.

In Panel (IV), it is possible to notice that as we move up in the socioeconomic scale the `contribution' of the \ST\ covariate, in the range from 1 to 4 minimum wages, is negative, even though the there is an increasing trend. For land lots located in neighborhoods that correspond to more than 4 minimum wages, the effect is always positive; from 10 to 15 minimum wages the effect is uniform.

We note from Panel (V) that, contrary to what one would expect, the `contribution' of the \UC\ covariate is not positive. In the range from 3.0 to 5.0, the fitted curve displays small oscillations, alternating in the postive and negative regions. The positive effect only holds for utilization coefficients greater than 5.0.

Notice from Panel (VI) that as the front land lot (in logs) increases in highly priced neighborhoods the `contribution' of the log(\FRVN) covariate is mostly increasing and positive. However, in the 1.5 to 2.0 interval the positive effect is approximately uniform.



\subsection{Comparing models}

In order to compare the best estimated models via CNLRM (Model (1.4)), GLM (Model (2.1)) and GAMLSS (Model (3.5)) we shall use the AIC and BIC.\footnote{The criteria shall only be used to compare models that use the response (\UP) in the same measurement scale: Models (2.1) and (3.5).} We shall also compare the different models using the pseudo-$R^2$ given in 
\eqref{psdr}.

We present in Table \ref{tabelacriterio} a comparative summary of the three models. We note that Model (3.5) is superior to the two competing models. Not only it has the smallest AIC and BIC values (in comparison to Model (2.1)), but it also has a much larger pesudo-$R^2$. The GAMLSS pseudo-$R^2$ exceeds 0.80, which is notable. 

\begin{table}[htb!]
\scriptsize
\begin{center}
	\caption{Comparative summary of the CNLRM, GLM and GAMLSS estimated models.}\label{tabelacriterio}
	\begin{tabular}{rllrr}
  \hline
 Model &Class & AIC & BIC & Pseudo-$R^2$ \\ 
  \hline
(1.4) &(CNLRM) & 4290 & 4392 & 0.667 \\ 
(2.1) &(GLM) & 19486 & 19581 & 0.672 \\ 
(3.5) &(GAMLSS) & 18822 & 19212 & 0.811 \\ 
     \hline
\end{tabular}
\end{center}
\end{table}

\subsection{Dispersion parameter modeling ($\sigma)$}

After a suitable model for the prediction of $\mu$ was selected, we carried out a likelihood ratio test to determine whether the GAMLSS scale parameter $\sigma$ is constant for all observations. The null hypothesis that $\sigma$ is constant was rejected at the usual nominal levels. We then built a regression model for such a parameter. 
To that end, we used stepwise covariate selection, considered different link functions (such as identity, inverse, reciprocal, etc.) and included smoothing functions (such as cubic splines, loess and penalized splines) in the linear predictor, just as we had done for the location parameter. We used the AIC for selecting the smoothers and for choosing the number of degrees of freedom of the smoothing functions together with visual inspection of the smoothed curves.

We present in Table \ref{ajustemod36} the GAMLSS hedonic price function parameter estimates obtained by jointly modeling the location $(\mu)$ and dispersion $(\sigma)$ effects; Model (3.6). The model uses the gamma distribution for the response and the log link function for both $\mu$ and $\sigma$. We note that Model (3.6) contains parametric and nonparametric terms, and for that reason it is said to be a linear additive semiparametric GAMLSS.

 \begin{table}[!ht]
 \scriptsize
\begin{center}
\caption{Hedonic price function estimated via GAMLSS -- Model (3.6).}\label{ajustemod36}
\begin{tabular}{rrrrr}
  \hline\hline
   $\mu$ Coefficients&&&&\\
  \hline\hline
&Estimative  &Standard error & $z$-statistic  & $p$-value\\ 
  \hline
(Intercept) & $-95.1300$ & 14.2700 & $-6.665$ & 0.0000 \\ 
  cs(LAT, df=10) & 5.94e-05 & 5.37e-06 & 11.053 & 0.0000 \\ 
  cs(LON, df=10) & 6.45e-06 & 1.86e-06 & 3.460 & 0.0000 \\ 
  cs(log(AR), df=10) & $-0.2087$ & 0.0104 & $-20.138$ & 0.0000 \\ 
  cs(ST, df=8) & 0.0321 & 0.0030 & 10.666 & 0.0000 \\ 
  cs(UC, df=3) & 0.2095 & 0.0161 & 13.006 & 0.0000 \\ 
  STR1 & 0.2039 & 0.0298 & 6.838 & 0.0000 \\ 
  STR2 & 0.0729 & 0.0276 & 2.635 & 0.0084 \\ 
  SI & 0.7136 & 0.0192 & 3.705 & 0.0000 \\ 
  PA & 0.1653 & 0.0157 & 10.465 & 0.0000 \\ 
  TO & 0.1778 & 0.0370 & 4.799 & 0.0000 \\ 
  NIO & 0.3722 & 0.0251 & 14.799 & 0.0000 \\ 
  NIT & 0.2790 & 0.0468 & 5.957 & 0.0000 \\ 
  YR06 & 0.1255 & 0.0175 & 7.144 & 0.0000 \\ 
  YR07 & 0.4195 & 0.0177 & 23.622 & 0.00 \\ 
  cs(log(FRVN), df=10) & 0.6809 & 0.0403 & 16.88 & 0.0000 \\ 
  SZ & 0.4824 & 0.0241 & 20.001 & 0.0000 \\ 
  \hline\hline
  $\sigma$ Coefficients&&&&\\
  \hline\hline
  (Intercept) & $-1.6838$ & 0.0839 & $-20.072$ & 0.0000 \\ 
  cs(log(AR), df=10) & 0.1370 & 0.0143 & 9.593 & 0.0000 \\ 
  ST & $-0.0391$ & 0.0040 & $-9.632$ & 0.0000 \\ 
   \hline
\end{tabular}
\end{center}
\end{table}

\begin{figure}[htbp]
	\centering
		\includegraphics[scale=0.38]{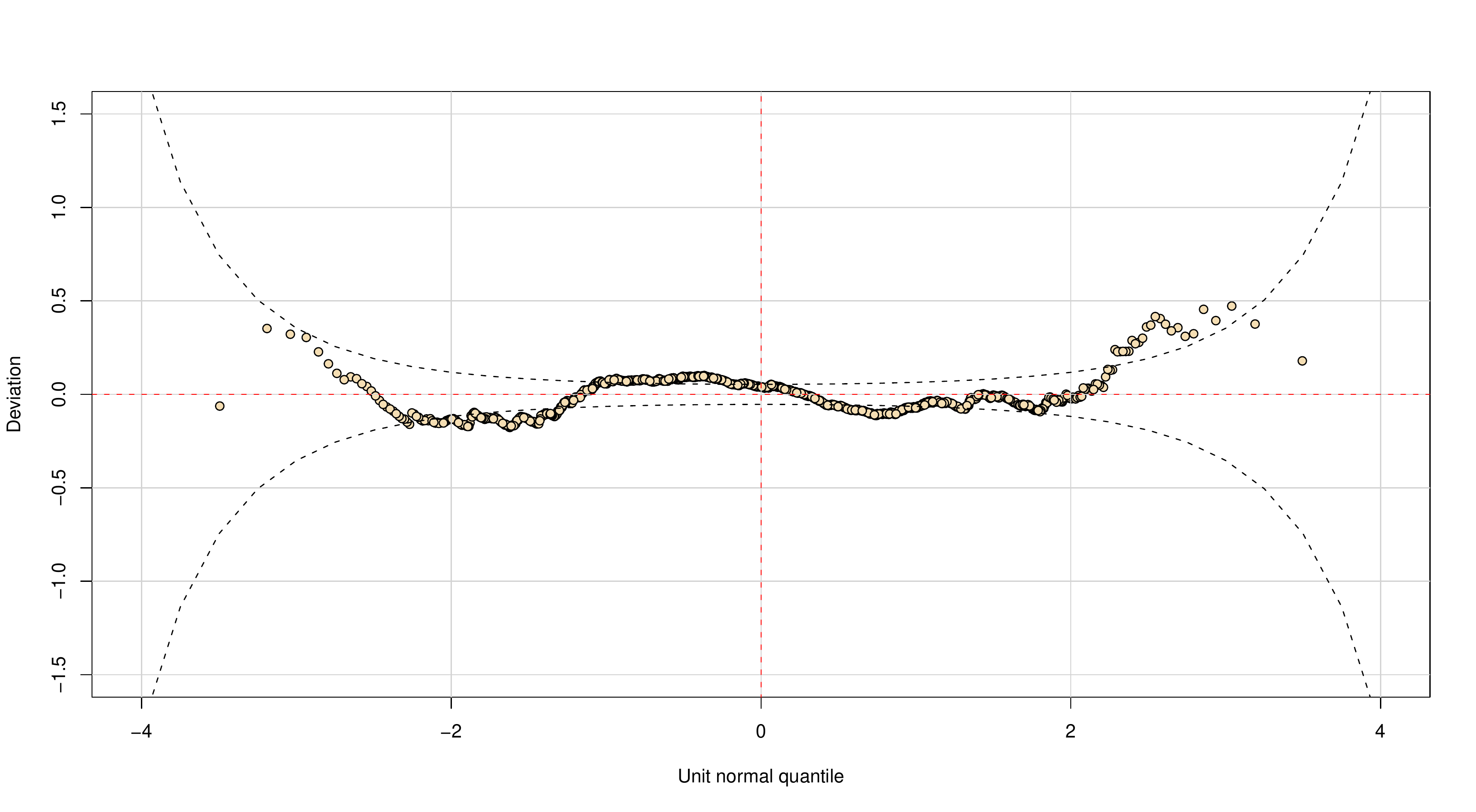}
	\caption{Worm plot -- Model (3.5).}\label{figwormmod35GAMLSS}
	\vskip-0.4cm
	\end{figure}

\begin{figure}[htbp]
	\centering
		\includegraphics[scale=0.38]{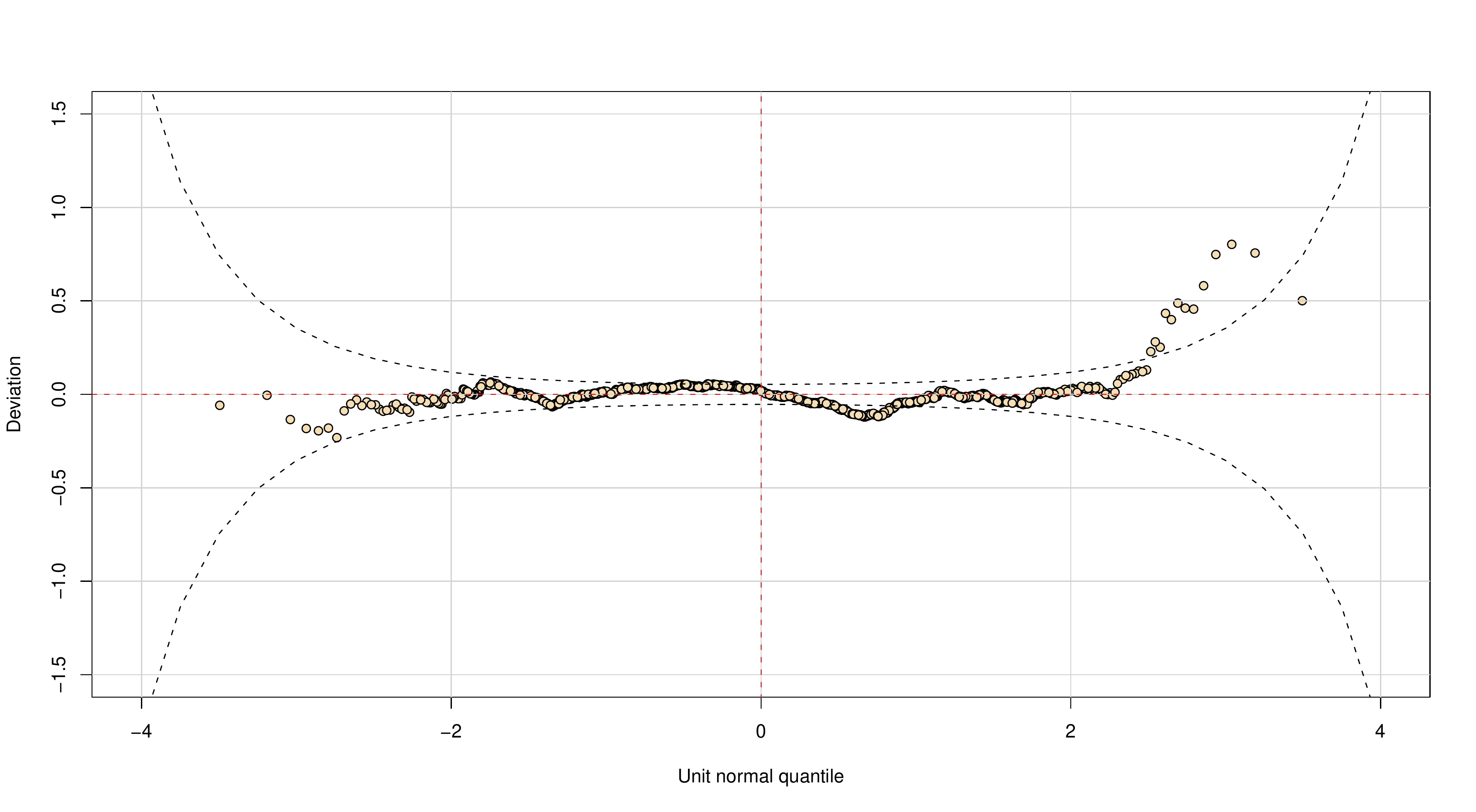}
	\caption{Worm plot -- Model (3.6).}\label{figwormmod36GAMLSS}
	\end{figure}

We note from Table \ref{ajustemod36} that the parameter estimates of the location submodel in Model (3.6) are similar to the corresponding estimates from Model (3.5), in which $\sigma$ was taken to be constant; see Table~\ref{ajustemod35}. It is noteworthy, nonetheless, that there was a sizeable reduction in the AIC, BIC and GD values (18607, 19065 and 18445, respectively) and also an improvement in the residuals as evidenced by the worm plot; see Figures \ref{figwormmod35GAMLSS} and \ref{figwormmod36GAMLSS}.

Only two covariates were selected for the $\sigma$ regression submodel in Model (3.6), namely: \ST\ and log(\AR). The former (\ST) entered the model in the usual parametric fashion whereas the latter (log(\AR)) entered the model nonparametrically through a cubic spline smoothing function with ten effective degrees of freedom. We note that the positive sign of the log(\AR) coefficient indicates that the \UP\ dispersion is larger for land lots with larger areas whereas the negative sign of the \ST\ coefficient indicates that the dispersion is inversely related to the socioeconomic neighborhood indicator. 


\begin{figure}[htb!]
	\centering
		\includegraphics[height=8.0cm,width=14.0cm]{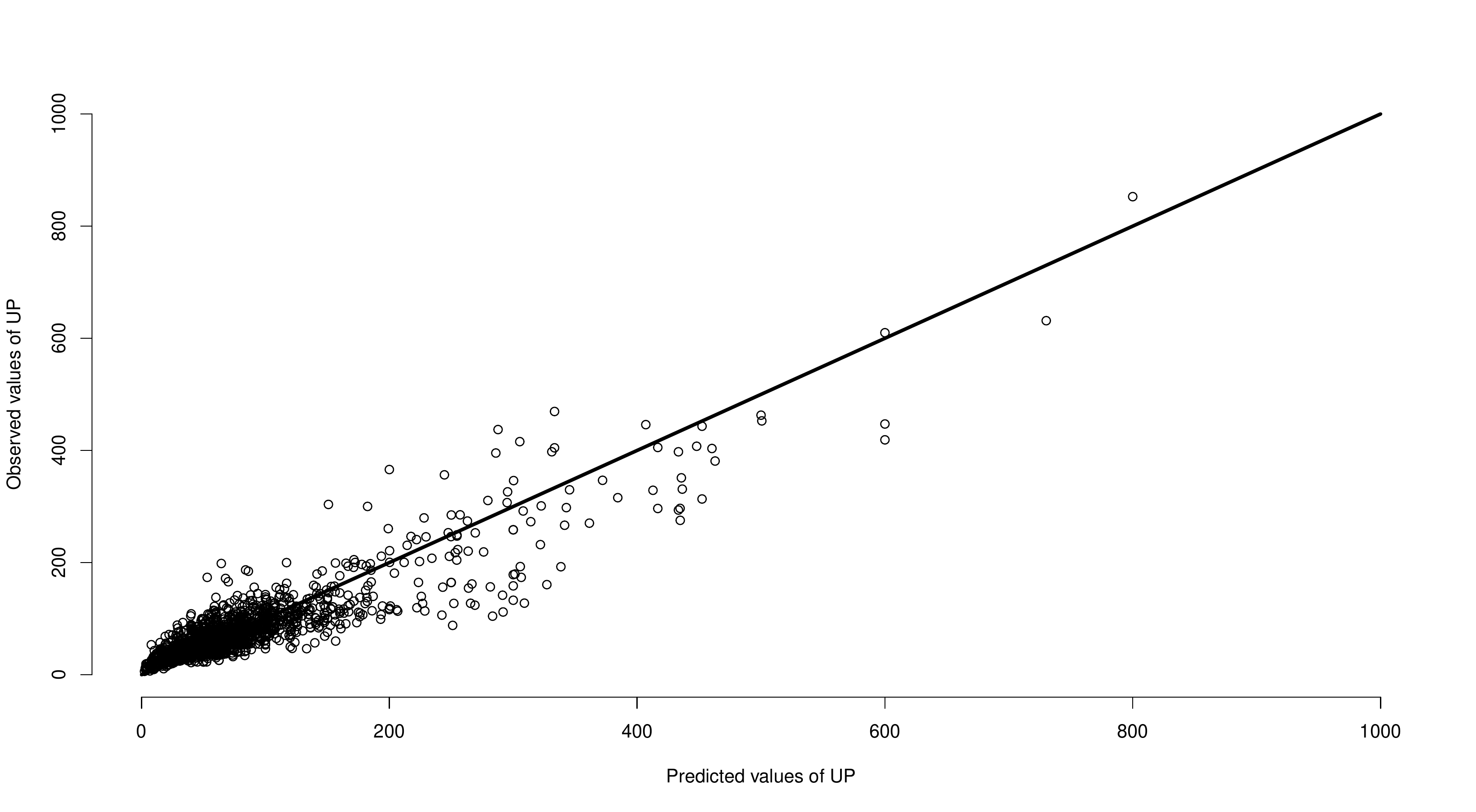}
	\caption{Observed values $\times$ predicted values of \UP\ -- Model (3.6).}\label{figvpmod36GAMLSS}
	\end{figure}

It is noteworthy that the pseudo-$R^2$ of Model (3.6) is quite high (0.817) and that all of explanatory variables are statistically significant at the 1\% nominal level which is not all that common in large sample cross sectional analyses, especially in real estate appraisals. Overall, the variable dispersion GAMLSS model is clearly superior to the alternative models. The good fit of Model (3.6) can be seen in Figure \ref{figvpmod36GAMLSS} where  we plot the observed response values against the predicted values from the estimated model. Note that the $45^{\mathrm{o}}$ line in this plot indicates perfect agreement between predicted and observed values. 

Model (3.6) is given by 
\begin{align*}
\log(\mu)&=\beta_0+\mathrm{cs}(\textrm{LAT},\: \textrm{df}=10)+\mathrm{cs}(\textrm{LON},\: \textrm{df}=10)+\mathrm{cs}(\textrm{log(AR),\: \textrm{df}}=10) + \\
 &\mathrm{cs}(\textrm{UC},\: \textrm{df}=3)+\mathrm{cs}(\textrm{ST},\: \textrm{df}=8)+\beta_1\textrm{STR1}+\beta_2\textrm{STR2} + \beta_3\textrm{SI}+\\ &\beta_4\textrm{PA}+
\beta_{5}\textrm{TO}+\beta_{6}\textrm{NIO}+\beta_{7}\textrm{NIT}+\beta_{8}\textrm{YR06}+\beta_{9}\textrm{YR07}+\beta_{10}\textrm{SZ} +\\
&\mathrm{cs}(\textrm{log(FRVN)},\: \textrm{df}=10),\\
\log(\sigma) &= \gamma_0+\gamma_1\textrm{ST}+\mathrm{cs}(\textrm{log(AR)}, \textrm{df}=10),\\
\end{align*}
in which the response (\UP) follows a gamma distribution (GA) with location and scale parameters $\mu$ and $\sigma$, respectively. This model proved to be the best model for hedonic prices equation estimation of urban land lots in Aracaju.

\section{Concluding remarks}\label{S:5}

Real state appraisal is usually performed using the standard linear regression model or the class of generalized linear models. In this paper, we introduced real state appraisal based on the class of generalized additive models for location, scale and shape, GAMLSS. Such a class of regression models provides a flexible framework for the estimation of hedonic price functions. It even allows for some conditioning variables to enter the model in a nonparametric fashion. The model also accomodates variable dispersion and can be based on a wide range of response distributions. Our empirical analysis was carried out using a large sample of land lots located in the city of Aracaju (Brazil). The selected GAMLSS model displayed a very high pseudo-$R^2$ (approximately 0.82) and yielded an excellent fit. Moreover, the inclusion of nonparametric additive terms in the model allowed for the estimation of the hedonic price function in a very flexible way. We showed that the GAMLSS fit was clearly superior to those based on the standard linear regression and on a generalized linear model. We strongly recommend the use of GAMLSS models for real state appraisal.  


\section*{Acknowledgements}

LF acknowledges funding from Coordena\c{c}\~{a}o de Aperfei\c{c}oamento de Pessoal de N\'{\i}vel Superior (CAPES), FCN and RO acknowledge funding from Conselho Nacional de Desenvolvimento Cient\'{\i}fico (CNPq). We thank three anonymous referees for their comments and suggestions.

\end{document}